\documentclass[11pt,cite,epsfig,psfrag]{article}
\usepackage[utf8]{inputenc}
\usepackage{geometry}
 \usepackage{graphicx}
 \usepackage{subfig}
  \usepackage{amsmath} 
  \graphicspath{{images33/}}
  \usepackage[usenames, dvipsnames]{color}
\usepackage{wrapfig}
\usepackage{boxedminipage}
\usepackage{enumerate,amsmath,amssymb,hhline,multirow,array}
\newcommand{\RNum}[1]{\uppercase\expandafter{\romannumeral #1\relax}}

\usepackage{romannum}
\usepackage{array}
\newcolumntype{M}[1]{>{\centering\arraybackslash}m{#1}}

\usepackage{fullpage}
\usepackage{amsmath}
\usepackage{amssymb}
\usepackage{graphicx}
\usepackage{mathrsfs}
\usepackage{wrapfig}
\usepackage{boxedminipage}
\usepackage{epsfig}
\usepackage{setspace}
\usepackage{float}
\usepackage{hyperref}
\usepackage{enumerate}
\usepackage{cite}
\usepackage[font=footnotesize,labelfont=rm]{caption}

\usepackage[numbers,sort&compress]{natbib}

\def\be{\begin{equation}}
\def\ee{\end{equation}}
\def\bea{\begin{eqnarray}}
\def\eea{\end{eqnarray}}

\numberwithin{equation}{section}

 \newcommand{\RN}[1]{%
   \textup{\uppercase\expandafter{\romannumeral#1}}%
 }
\begin{document}

\pagenumbering{arabic}

\vskip 2cm

\begin{center}
{\Large \bf A Note on Circular Geodesics and Phase Transitions of Black Holes}
\end{center}

\vskip .2cm

\vskip 1.2cm

\begin{center}
	{ {\bf Chandrasekhar Bhamidipati}\footnote{ chandrasekhar@iitbbs.ac.in}\\
		\vskip 0.1cm School of Basic Sciences\\ 
		\vskip 0.5cm
	{\bf	Shrohan Mohapatra} \footnote{ sm32@iitbbs.ac.in}\\ 
	\vskip 0.1cm
	School of Electrical Sciences
	}
\end{center}

\vskip 5mm 
\begin{center}{
Indian Institute of Technology Bhubaneswar \\ Jatni, Khurda, Odisha, 752050, India}
\end{center}

\vskip 1.2cm
\vskip 1.2cm
\centerline{\bf Abstract}
\noindent

The circular motion of charged test particles in the gravitational field of a Reissner-Nordstr\"{o}m black hole in Anti de Sitter space-time is investigated, using a set of independent parameters, such as charge Q, mass M and cosmological constant $\Lambda= -3/l^2$ of the space-time, and charge to mass ratio $\epsilon=q/m$ of the test particles. Classification of different spatial regions where circular motion is allowed, is presented, showing in particular, the presence of orbits at special limiting values, $M=4/\sqrt{6} Q$ and $l=6 Q$. Thermodynamically, these values are known to occur when the black hole is on the verge of a second order phase transition, there by, giving an interesting connection between thermodynamics and geodesics of black holes.  We also comment on the possibility of such a connection for black holes in flat spacetime in a box.


\newpage
\setcounter{footnote}{0}
\noindent

\baselineskip 15pt

\section{Introduction}

Possible connections involving the geodesic motion, quasinormal modes (QNMs) and phase transitions of black holes are quite interesting to study. The motion of test particles in the vicinity of a black hole is the first probe towards an understanding of the physics and geometry of gravitational objects in general relativity. Quasinormal modes carry information about the stability and relaxation of the black hole under perturbations and are entirely fixed by the structure of the background geometry and interesting connections with thermodynamics have been explored~\cite{Liu:2014gvf, Berti:2008xu, Jing:2008an}. On the other hand, an important characteristic of a black hole is its thermodynamic property~\cite{Bekenstein:1973ur}-\cite{Gibbons:1976ue}. It is well known that the heat capacity of Schwarzschild black holes in flat space-time is always negative and hence the black hole is thermodynamically unstable. Situation changes, once charge and cosmological constant are introduced~\cite{Hawking:1982dh}. The Reissner-Nordstrom black hole has a heat capacity which is negative in some region of parameter space and positive in other regions. Davies showed that phase transitions occur in black holes and a second order phase transition (SOPT) takes place when the black hole parameters follow the relation~\cite{Davies:1978mf,Davies:1989ey}:
\bea \label{davis}
\left(\frac{Q}{M}\right)^2 = \left(\frac{9}{4}\right)^2 \Lambda M^2 + \frac{3}{4}\, ,
\eea
where $M,Q$ are mass and charge of the black hole, respectively, and $\Lambda = -3/l^2$ is the cosmological constant and $l$ is the Anti de Sitter (AdS) length scale. Relation (\ref{davis}) occurs at the point where the heat capacity diverges. More recently, phase transitions of black holes in AdS space-times has been boosted by a modified approach of extended phase space formalism, where the cosmological constant is dynamical~\cite{Kastor:2009wy}-\cite{Johnson:2018amj}, giving rise to pressure $p=3/{8\pi l^2}$. Various aspects of black holes such as, complete mapping to Van der Waals transition, P-V criticality and novel results in holography have emerged. The thermodynamic phase structure of black holes includes a line of first order phase transitions ending in a second order transition point~\cite{Chamblin,Chamblin:1999hg,Kubiznak,Johnson:2017asf}. At the critical point, the thermodynamic quantities take specific values scaling with respect to charge $Q$ as:
\begin{equation}
\label{criticalpoint}
r_{\rm cr}=\sqrt{6}Q\ ,\quad T_{\rm cr}= \frac{1}{3\sqrt{6}\pi Q}\ ,  \quad p_{\rm cr}=\frac{1}{96\pi Q^2}\ , 
\end{equation}
where entropy and thermodynamic volume scale as $S_{\rm cr}=6\pi Q^2$  and $V_{\rm cr}= 8\sqrt{6}\pi Q^3$. Here, $r_{\rm cr}, T_{\rm cr}$ and $ p_{\rm cr}$ are the critical values taken by horizon radius, temperature and pressure associated with the black hole, respectively. Critical values taken by other parameters
\begin{eqnarray}\label{criticalAdS}
&& l_{\rm cr}=6 \, Q \, \, M_{\rm cr}=\frac{4}{\sqrt{6}}\, Q \, ,
\end{eqnarray}
satisfy the relation given by Davis in eqn. (\ref{davis}). Instead of the perturbations from external modes of a field (as in QNM study), if we specialise to test particle motion in black hole background in the probe limit, a related question is whether there is a connection between the study of circular geodesics (which also has to do with the fixed geometry of the background space-time) and thermodynamics of black holes. An interesting advance in this direction was recently made in~\cite{Wei:2017mwc}, where the authors showed that the radius and impact factor corresponding to circular geodesics carries information about the first order phase transition of black hole in AdS. In this note, our aim is to study the question whether the circular geodesics carry information about the second order phase transitions of black holes, and we find that they do.\\

\noindent
Let us note that the geodesics of black holes, in general, have already been studied in ample detail in a variety of contexts~\cite{Chandrasekhar1983}-\cite{Pugliese:2013xfa}. It is well known that the stable and unstable orbits can be classified by drawing the effective potential and investigating its behaviour at various points. However, our interest is in studying the geodesics to look for points in the parameter space which capture the SOPT, in particular, satisfying relations such as the one given in eqn. (\ref{davis}). To this end, we require a classification of circular geodesics which depends only on a few independent parameters corresponding to the black hole. One particular line of study involving classification of circular geodesics in the background of charged black holes in flat space-time (with no cosmological constant), presented in~\cite{Pugliese:2013xfa} is quite relevant for this problem 
In this reference, the authors performed a classification of different spatial regions around the black hole where circular orbits exist, based on a few independent parameters depending on the black hole (such as, Charge $Q$ and Mass $M$) and the test particle charge to mass ratio $\epsilon = q/\mu$. Among other things, the authors found the behaviour of circular orbits varies depending on the special limiting values:
\bea \label{flat}
 M/Q=2, 5/\sqrt{13}, \sqrt{3/2} \, ,
\eea 
and $\epsilon = 1,2, M/Q$ for positive charges (see~\cite{Pugliese:2013xfa} for full table of classification and classification in naked singularities).  Apart from being interesting in its own right, such a classification can also lead to distinction of black holes and naked singularities. Naivly, these values do not seem to have a connection with thermodynamics of the black hole. For example, setting the cosmological constant to zero in eqn. (\ref{davis}), gives the value $M/Q = 2/\sqrt{3}$, which does not match with the locations of the circular orbits in eqn. (\ref{flat}). We will comment on this issue later in section-3.  \\

\noindent
We now summarise our findings when the cosmological constant is included. Keeping the relation in eqn. (\ref{davis}) in mind, we first seek to extend the classification of circular orbits done in~\cite{Pugliese:2013xfa} to the case of black holes in AdS, where a new independent parameter is available, namely, the cosmological constant $\Lambda$.  Here, one expects the behaviour of circular orbits to depend on limiting values obtained not just from the independent parameter $M/Q$, but also a new dimensionless parameter $l/Q$. Generalising the analysis in~\cite{Pugliese:2013xfa} for AdS, we find several ranges for the independent parameters where circular motion is possible, giving special limiting values, which are summarised in Table-3. The main result of this letter is that the classification of orbits around the black holes in AdS throws open a new value where circular motion is possible, namely: $l=6 Q$ and $M=4/\sqrt{6} Q$ (See the equation coming from thermodynamics study in (\ref{criticalAdS}). Unlike other values of parameters in Table-3, this is special, because the black hole is known to undergo a second order phase transition precisely when the parameters take these values~\cite{Chamblin,Chamblin:1999hg,Kubiznak,Johnson:2017asf}. Moreover, this set of values obtained from the study of geodesics, also satisfies the relation obtained by Davis from thermodynamics, given in eqn. (\ref{davis}).  \\

\noindent
Rest of this note is organised as follows. In Section-(2.1), we set up the main equations obtaining angular momentum and energy, necessary for studying circular geodesics of charged test particles in black holes in AdS. Section-(2.2) contains our main results on the classification of geodesics in AdS and connection to critical point of phase transitions. In Section-3, we end with remarks on extension of results of section-(2.2) to black holes in flat spacetime in a box.

\section{Motion of Test Particles in Black Holes in AdS}
\subsection{Circular Geodesics of Charged Particles in AdS}

We start with the action for Einstein-Maxwell system in the presence of a negative cosmological constant $\Lambda=-3/l^2$, given as:\footnote{Newton's constant $G$, speed of light $c$ and Planck's constant $\hbar$ are set to unity.}
\begin{equation}
I=-\frac{1}{16\pi }\int \! d^4x \sqrt{-g} \left(R-2\Lambda -F^2\right)\ .
\label{eq:action}
\end{equation}
The black hole space-time is Reissner--Nordstr\"om (RN) like solution with metric
\begin{eqnarray} \label{metric}
ds^2 &=& -Y(r)dt^2
+ {dr^2\over Y(r)} + r^2 d\Omega_{2}^2\ , \\
Y(r) &\equiv& 1-\frac{2 M}{r}+\frac{Q^2}{r^{2}}+\frac{r^2}{l^2}\ , \quad A_t = Q\left(\frac{1}{r_+}-\frac{1}{r}\right)\, ,
\end{eqnarray}
where $d\Omega_{2}$ is the metric on $S^{2}$ and $A_t$ is the gauge potential. 
The motion of a test particle of charge $q$ and mass $\mu$ moving in a
RN  background (\ref{metric}) is described by the
following Lagrangian density and equations of motion, respectively:
\begin{eqnarray}\label{LRNADS}
&&\mathcal{L}=\frac{1}{2} g_{\alpha\beta}\dot{x}^{\alpha}\dot{x}^{\beta}+\epsilon A_\alpha x^\alpha, \nonumber \\
&&\dot{x}^{\alpha}\nabla_{\alpha}\dot{x}^{\beta}=\epsilon F^{\beta}_{\ \gamma}\dot{x}^{\gamma},
\end{eqnarray}
where $\epsilon = q/\mu$ is the specific charge of the test particle and
$F_{\alpha\beta}\equiv A_{\alpha,\beta}-A_{\beta,\alpha}$. Since the Lagrangian  density (\ref{LRNADS})  does not depend explicitly on the variables $t$ and $\phi$, the following two conserved quantities exist
\begin{eqnarray}
\label{pt} p_t&\equiv& \frac{\partial\mathcal{L}}{\partial
	\dot{t}}=-\left(Y(r)\dot{t}+\frac{\epsilon Q}{r}\right)=-\frac{E}{\mu},\\
\label{pphi}  p_{\phi}&=&\frac{\partial\mathcal{L}}{\partial
	\dot{\phi}}=r^2\sin^2\theta \dot{\phi}=\frac{L}{\mu},
\end{eqnarray}
where $L$ and  $E$ are respectively the angular momentum and energy
of the particle as measured by an observer at rest at infinity. We restrict ourselves to the study of equatorial trajectories with $\theta =\pi/2$ where,
\begin{equation} \label{oneDequiv}
\dot r^2 + V^2 = E^2/\mu^2 \, ,
\end{equation}
which describes the motion inside an effective potential given as:
\begin{equation}\label{veff}
V_{\pm}=\frac{E^{\pm}}{\mu}=\frac{\epsilon Q}{r}\pm
\sqrt{\left(1+\frac{L^2}{\mu^2r^2}\right)\left(1-\frac{2M}{r}+\frac{Q^2}{r^2} +\frac{r^2}{l^2} \right)} \, .
\end{equation}
At the turning point, we set $(V=E/\mu)$, the point where the kinetic energy of the particle vanishes. 
Considering the special case of $V_+$, as in~\cite{Pugliese:2010ps,Pugliese:2010he}, conditions for circular orbits are \footnote{Subscript $(+)$ will be dropped henceforth}
\begin{equation}\label{fg2}
\frac{d V}{d r}=0, \quad V=\frac{E}{\mu}.
\end{equation}
Angular momentum is found to be:
\begin{eqnarray}\label{ang1}
&&\frac{\left(L_{\pm}^2\right)}{\mu^2} =  \frac{1}{2 l^2 \left(r (r-3 M)+2 Q^2\right)^2}\left( r^6 \left(2 r (r-3 M)+Q^2 \left(\epsilon ^2+4\right)\right) -l^2\, A \right)\, ,
\end{eqnarray}
where $A = \left( \pm \sqrt{A_1}+ A_2 \right)$ and
\begin{eqnarray}
&& A_1 = \frac{Q^2 r^4 \epsilon ^2 \left(l^2
	\left(-2 M r+Q^2+r^2\right)+r^4\right)^2 \left(4 r (r-3 M)+Q^2 \left(\epsilon ^2+8\right)\right)}{l^4} \, ,\nonumber \\
&& A_2 = 6 M^2
r^4+Q^2 r^3 \left(2 M \left(\epsilon ^2-5\right)-r \left(\epsilon ^2-2\right)\right)-2 M r^5-Q^4 r^2
\left(\epsilon ^2-4\right) \, .
\end{eqnarray}
Energy is obtained as
\begin{eqnarray}
\label{ener1}
&& E= \frac{1}{2 r}\left(\sqrt{2} r \sqrt{\frac{\left(l^2 \left(-2 M r+Q^2+r^2\right)+r^4\right)}{l^4 r^4 \left(r (r-3 M)+2 Q^2\right)^2}		
	\left(	\left(l^2 \left(\sqrt{A_1}+12 M^2 r^4+Q^2 r^3 \left(r \left(\epsilon ^2+6\right)-2 M \left(\epsilon
			^2+7\right)\right) \right.\right. \right. }.\right. \nonumber \\
&& \left. {.\left.\left.\left. -10 M r^5+Q^4 r^2 \left(\epsilon ^2+4\right)+2 r^6\right)+r^6 \left(2 r (r-3 M)+Q^2
			\left(\epsilon ^2+4\right)\right)\right) \right)}		
		+2 Q \epsilon \right) 
\end{eqnarray}
The above expressions in equations (\ref{ang1}) and (\ref{ener1}) go over to the ones in~\cite{Pugliese:2013xfa} as the cosmological constant is set to zero.
The radius of an orbit where particle is located at rest as seen by an observer at infinity, i.e., $L=0,  \frac{d V}{dr}=0$ and the radius of the last stable circular orbit can be computed numerically, but we do not pursue it here.

\subsection{Classification of Circular Orbits and Critical Point of Phase Transition}

We now investigate the conditions for the existence of circular orbits of charged test particles in the background of charged black holes in AdS, with the metric given in eqn. (\ref{metric}). Instead of the general approach of using the effective potential,  we follow the alternate methodology of~\cite{Pugliese:2013xfa}, where we first express the conditions for circular motion in terms of the independent parameters $M/Q$, $l/Q$ and the charge to mass ratio of the test particle $\epsilon=q/\mu$.  Unlike the  flat space computation performed in~\cite{Pugliese:2013xfa}, where only one independent parameter $M/Q$ was available, in the present case, we also have the parameter $l/Q$ on which the orbits can depend. The physical values of the parameters are obtained from the conditions for existence of circular motion, i.e., positivity of physical quantities such as angular momentum and energy of the charged particles $(k=1)$ given in equations (\ref{ang1}) and (\ref{ener1}), respectively.\\

\begin{table}[ht!]
\caption{Values of the angular momentum in different ranges of radii.}
\begin{tabular}{|c|c|c|c|}
\cline{1-3}
Region & Range of radii & Value of Angular Momentum \\ \cline{1-3}
\cline{1-3}
A & $r \in [r_{\epsilon}^{+},\infty)$ & $L_{\pm}$ \\ \cline{1-3}
B & $r \in (r_h,\infty)$ & $L_{\pm}$ \\ \cline{1-3}
C & $r \in (r_{\epsilon}^{+},\infty)$ & $L_{\pm}$ \\ \cline{1-3}
D & $r \in (r_{\gamma}^{+},\infty)$ & $L_{\pm}$ \\ \cline{1-3}
\end{tabular}
\end{table}
\begin{table}[ht!]
\caption{Classes (I,II,III) based on charge to mass ratio $\epsilon$ and the ratio $\frac{M}{Q}$ of the RN AdS black hole.}
\begin{tabular}{|M{4cm}|M{1cm}|M{4cm}|M{1cm}|M{2cm}|M{1cm}|}
\hline
\multicolumn{2}{|c|}{\Romannum{1}} & \multicolumn{2}{c|}{\Romannum{2}} & \multicolumn{2}{c|}{\Romannum{3}} \\ \hline
Range & Region & Range & Region & Range & Region \\ \hline
$\epsilon = 0$ & D & $\epsilon = 0$ & D & $\epsilon = 0$ & D \\ \hline
$0 < \epsilon < \epsilon_{h}$ & C & $0 < \epsilon < \sqrt{9\frac{M^2}{Q^2}-8}$ & C & $0 < \epsilon < \epsilon_{h}$ & C \\[25pt] \hline
$\epsilon_{h} \le \epsilon < \sqrt{9\frac{M^2}{Q^2}-8}$ & A & $\epsilon = \sqrt{9\frac{M^2}{Q^2}-8}$ & B & $\epsilon \ge \epsilon_{h}$ & B \\[25pt] \hline
$\epsilon > \sqrt{9\frac{M^2}{Q^2}-8}$ $\epsilon = \sqrt{9\frac{M^2}{Q^2}-8}$ & D\hskip 0.5cm B & $\epsilon > \sqrt{9\frac{M^2}{Q^2}-8}$ & D  & &  \\[25pt] \hline
\end{tabular}
\end{table}
\begin{table}[ht!]
\caption{Possible values of $\frac{M}{Q}$ for a given $\frac{l}{Q}$ parameter, of the RN AdS black hole}
\begin{tabular}{|M{1.5cm}|c|M{1.5cm}|c|M{1.5cm}|c|M{1.75cm}|c|M{1.5cm}|c|}
\hline
\multicolumn{2}{|c|}{$\frac{l}{Q}\in(0,2)$} & \multicolumn{2}{c|}{$\frac{l}{Q}=2$} & \multicolumn{2}{c|}{$\frac{l}{Q}\in(2,6)$} & \multicolumn{2}{c|}{$\frac{l}{Q}=6$} & \multicolumn{2}{c|}{$\frac{l}{Q}\in(6,\infty)$} \\ \hline
Range & Class & Range & Class & Range & Class & Range & Class & Range & Class \\ \hline
$\frac{M}{Q} \in (r_l,\infty)$ & \Romannum{1} & $\frac{M}{Q} \in (\frac{4}{3}\sqrt{\frac{2}{3}},\infty)$ & \Romannum{1} & $\frac{M}{Q} \in (r_l,\infty)$ & \Romannum{1} & $\frac{M}{Q} \in (\frac{4}{3}\frac{1}{3^{\frac{1}{4}}},2\sqrt{\frac{2}{3}})$ & \Romannum{1} & $\frac{M}{Q} \in (r_l,r_l^{+})$ & \Romannum{1} \\ \hline
 & & & & & & $\frac{M}{Q} = 2\sqrt{\frac{2}{3}}$ & \Romannum{2} & $\frac{M}{Q} = r_l^{+}$ & \Romannum{2} \\ \hline
 & & & & & & $\frac{M}{Q} \in (2\sqrt{\frac{2}{3}},\infty)$ & \Romannum{1} & $\frac{M}{Q} \in (r_l^{+},r_l^{-})$ & \Romannum{3} \\ \hline
 & & & & & & & & $\frac{M}{Q} = r_l^{-}$ & \Romannum{2} \\ \hline
 & & & & & & & & $\frac{M}{Q} \in (r_l^{-},\infty)$ & \Romannum{1} \\ \hline 
\end{tabular}
\end{table}
\noindent
Demanding the positivity of Energy and Angular momentum gives rise to the following parameters:
\begin{equation}\label{one}
r_{\epsilon}^{\pm} = \frac{1}{2}(3 M \pm \sqrt{9 M^2 - 8 Q^2 - \epsilon^2 Q^2}) \,,
\end{equation}
\begin{equation}\label{two}
r_{\gamma}^{\pm} = \frac{1}{2}(3 M \pm \sqrt{9 M^2 - 8 Q^2}) \,,
\end{equation}
which are same as the flat space expressions\footnote{except for minor relableling suitable in present context} discussed in~\cite{Pugliese:2013xfa}.   Here, $r_{\gamma}^{\pm}$ and $r_{\epsilon}^{\pm} $ are the limiting radii at which neutral (photons) and charged test particles, respectively, can be in a circular orbit around the black hole. These radii appear in Table-1. 
In the presence of a cosmological constant, there are additional new parameters:
\begin{equation}
r_l =\frac{1}{ \sqrt{54}} \sqrt{36-\left(\frac{l}{Q}\right)^2+\frac{Q}{l}\left(12-\left(\frac{l}{Q}\right)^2\right)^\frac{3}{2}}\, ,
\end{equation}
\begin{equation}
r_l^{\pm} = \frac{1}{3}\sqrt{\frac{2}{3}} \sqrt{\pm\sqrt{\left(\frac{l}{Q}\right)^2 \left(\left(\frac{l}{Q}\right)^2 - 36 \right)} + \left(\frac{l}{Q}\right)^2}\, .
\end{equation}
The new parameters $r_l$ and $r_l^{\pm}$ appear in AdS space time, in the presence of cosmological constant and aid in the classification analysis, playing the role of limiting radii for circular motion in certain ranges (Table-3). The physical meaning of these radii is less clear.\\

\noindent
The classification of spatial regions where circular motion is possible, is divided in to Tables-1,2 and 3. Let us start from Table-3. Table-3 contains all possible limiting values of the ratio $M/Q$, for a given limiting value of the ratio $l/Q$. The possible limiting values of $M/Q$ fall in classes I, II and III. Every class is further sub divided in Table-2, in to certain regions denoted as $A,B,C,D$, based on the range of values taken by the charge to mass ratio $\epsilon$, in terms of the ratios $M/Q$ and $l/Q$.  For a given value of $\epsilon$ (in a certain class and region) in Table-2, the corresponding possible values of orbit radius and anglular momentum where circular motion is possible, are finally noted in Table-1. The appearance of the limiting radius $r_h$  in Table-1, given as $r_h = 3M/2$, is quite interesting, as it appears independently in thermodynamic description.  It is the radius of the black hole horizon at the phase transition point~\cite{Davies:1989ey}. The value of $\epsilon_h$, the limiting charge to mass ratio is given in Appendix A. \\

\noindent
Let us note our main result based on Table-3. The point in the parameter space where circular orbits exist, with value $l/Q=6$ and $M/Q = 2 \sqrt{\frac{2}{3}}$, exactly corresponds to the point where the black hole is critical and undergoes second order phase transition. 
A possible explanation for this connection is probably scale invariance at the critical point of black hole phase transition, which is a special point where all the thermodynamic quantities scale with respect to charge ${Q}$. The method used for classification of circular geodesics in this note is well suited to bring out the scaling of various parameters with the charge $Q$ of the black hole. Thermodynamically, the scaling implies that, Entropy $S\sim {Q}^2$, Pressure $p\sim {Q}^{-2}$, and Temperature $T\sim {Q}^{-1}$. There are important consequences for the geometry of the black hole, which itself now depend on a single parameter ${Q}$, which can be tuned and taken to be large. In fact, a new double scaling limit can be taken, where the charge is taken to be large while at the same time nearing the horizon, the geometry of the black hole turns out to be a fully decoupled Rindler space-time, much like the $AdS_2 \times S^2$ space-time obtained from the near horizon limit of extremal black holes~\cite{Johnson:2017asf}.  The connection of circular geodesics with phase transitions noted here is quite intriguing and needs further study, which might
might give a novel arena to explore the thermodynamics of black holes in AdS and holography, from the motion of test particles. On the other hand, let us also note here that the other limiting values in Table-3 where circular orbits exist, for instance $l/Q = 2, M/Q =4\sqrt{2}/{3\sqrt{3}}$ etc., do not seem to have any obvious analogue in thermodynamics and they also do not satisfy the Davis relation in eqn.(\ref{davis}). This may also mean that the above found connection between circular orbits and phase transitions of black holes might just be a nice coincidence (as it occurs only for single limiting value coming from the study of circular orbits, and not at other values where also circular orbits exist).

\section{Remarks on Charged Black Holes in Flat Spacetime in a Box}

In this paper, we presented a classification of circular orbits of charged particles in the background of Reissner-Nordstrom black holes in AdS, using a few independent parameters based on charge $Q$, mass $M$ and cosmological constant $l$, together with the charge-to-mass ratio $\epsilon$ of test particles. One particular limiting value taken by these parameters exactly coincides with the point coming independently from thermodynamic considerations on the black hole side, namely the critical point of a phase transition:  $l=6 Q$ and $M=4/\sqrt{6} Q$. We also noted that when the cosmological constant is set to zero, such a connection between location of circular orbits and phase transition of black holes ceases to exist, for example in flat space-time. The location of circular orbits as obtained in~\cite{Pugliese:2013xfa} and given in eqn. (\ref{flat}), does not satisfy the relation in eqn. (\ref{davis}). A possible reason for non-existence of such a connection in flat space-time may be understood from that the fact that thermodynamics of black holes in asymptotically flat spacetime does not meet the requirements of stability due to Hawking radiation. In order to remedy this situation, various earlier proposals exist~\cite{Whiting:1988qr,Braden:1990hw,York:1986it,Brown:1992br}. More recently, there has been a steady progress with respect to the study of black holes in a box~\cite{Carlip:2003ne}-\cite{Dias:2018zjg}. RN black holes in flat and AdS backgrounds are exact solutions of corresponding Einstein equations, with the main difference being the Hawking-Page transition seen in AdS. AdS background provides a natural box, due to the presence of a cosmological constant, allowing for a thermodynamic equilibrium between hot AdS gas and large black holes. There is currently lot of activity in  extended black hole thermodynamics, where the cosmological constant is treated as a dynamical variable, giving rise to pressure and a new $pdV$ term in the first law of black hole mechanics~\cite{Kastor:2009wy}-\cite{Johnson:2018amj}(and references there in). Analogous proposals for black holes in flat space time are emerging, where an artificial concentric cavity of a certain radius, say, $r_B$ is added, whose specific value is to be determined. There are also proposals, that the holographic features of black holes in AdS are possibly due to the natural confining box, rather than the specific details of the theory~\cite{Carlip:2003ne,Lundgren:2006kt,Ma:2016vop}. It has been noted in these works, that having the black hole in a box leads to a Hawking -Page transition and there also exists a second order phase transition, exactly as in AdS.\\

\begin{figure}[h]
	\begin{center}
		\centering
		\includegraphics[width=3.2in]{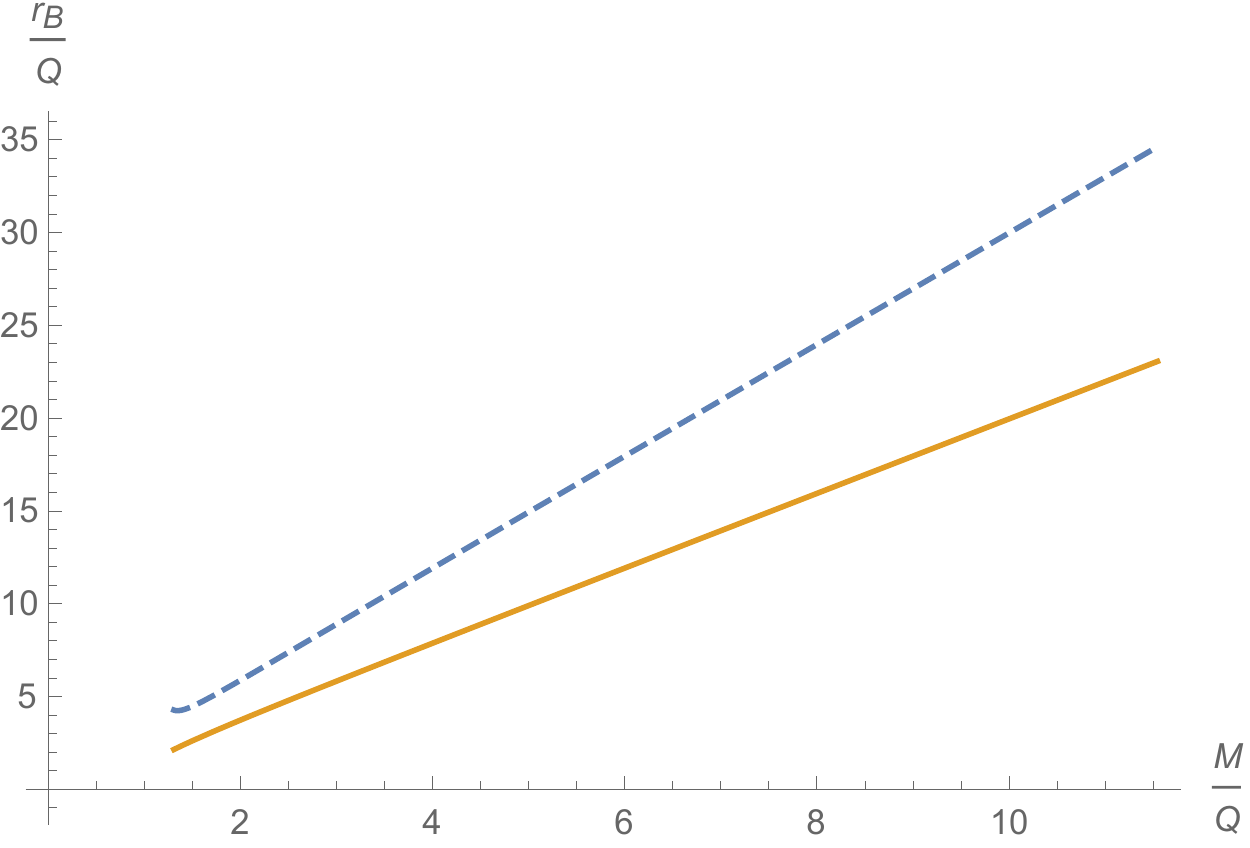}  		
		\caption{Dashed line corresponds to the thermodynamic phase boundary. Solid line shows the bound $r_b> r_+$. Specific heat is positive above the dashed line and negative below it.}   \label{rbQM}		
	\end{center}
\end{figure}

\noindent
Just like the Davis relation in eqn. (\ref{davis}), we now see that it is possible to obtain a relation among the thermodynamic parameters of flat space black holes in a box, by studying the divergence of specific heat. Basic set up of charged black holes in flat spacetime in a box of radius $r_B$ are discussed in detail in~\cite{Carlip:2003ne,Lundgren:2006kt,Ma:2016vop} and here, we only need to recall few important results. The metric for charged black holes in flat spacetime is given by
\begin{equation}\label{5}
 d{s^2}=-f(r)d{t^2} + \frac{{d{r^2}}}{{f(r)}} + {r^2}\left(d\theta^2 + \sin^2\theta\,d\phi^2 \right)\, ,
\end{equation}
where
\begin{equation}\label{6}
f(r) = 1 - \frac{{2M}}{r} + \frac{{{{Q}^2}}}{{{r^2}}}.
\end{equation}
To set up the thermodynamics,  black holes are enclosed in a cavity of radius $r_B$ with $r_B > r_+$, where $r_+$ is the outer horizon radius. The temperature is fixed at the surface of this cavity, to ensure that the system can be thermodynamically stable. To make contact with the study of circular geodesics, it is useful to also fix the charge in the cavity and study the system in the canonical ensemble. The action is known to be~\cite{Carlip:2003ne,Lundgren:2006kt,Ma:2016vop,Kubiznak:2016qmn}:
\begin{equation}\label{7}
\begin{aligned}
{I_E}&({r_B},{T_B},Q;{r_ + }) 
     ={\beta _B}{r_B}\left( {1 - \sqrt {\left( {1 - \frac{{{r_ + }}}{{{r_B}}}} \right)\left( {1 - \frac{{{{Q}^2}}}{{{r_B}{r_ + }}}} \right)} } \right) - \pi r_ + ^2 \, .
\end{aligned}
\end{equation}
Here $\beta_B=1/T_B$ and $T_B$ is the temperature of the cavity, given as:
\begin{equation}\label{13}
T_B =\frac{1}{4\pi r_ +}\frac{\left(1 - \frac{r_B Q^2}{r_ +} \right)^{1/2}}{\left(1 - \frac{r_B^2 Q^2}{r_ +^2}\right)r_ +\left(1 - {\frac{r_ +}{r_B}} \right)^{1/2}} \, .
\end{equation}
Using standard methods, specific heat in this case can be computed to be: 
\bea
C_p = \frac{4 S \left(\sqrt{S}-\sqrt{\pi } r_B\right) \left(S-\pi  Q^2\right) \left(r_B \sqrt{S}-\sqrt{\pi } Q^2\right)}{2 \sqrt{\pi } r_B^2
   \sqrt{S} \left(S-3 \pi  Q^2\right)+ r_B \left(5 \pi ^2 Q^4+6 \pi  Q^2 S-3 S^2\right)+2 \sqrt{\pi } Q^2 \sqrt{S} \left(S-3 \pi  Q^2\right)} \, .
\eea
The above expression diverges at:
\bea \label{diver}
\pi^2 (-6 Q^4 r_ + + 2 Q^2 r_ +^3 + r_B^2 (-6 Q^2 r_ + + 2 r_ +^3) + 
   r_B (5 Q^4 + 6 Q^2 r_ +^2 - 3 r_ +^4)) =0 \, ,
\eea
where we have eliminated entropy using $S=4\pi r^2$. As in~\cite{Davies:1989ey}, combining the above equation in (\ref{diver}) to the condition for existence of roots from the lapse function $f(r)=0$ from eqn. (\ref{6}), one finally gets the relation involving only thermodynamic parameters:
\begin{eqnarray} \label{davisflat} 
&& M \left(3 r_B^4 M-2 r_B^3 \left(7 M^2-9 Q^2\right) + 3 r_B^2 \left(5 M^3-4 M Q^2\right) \right.  \\ \nonumber 
 && \left. -2 r_B \left(7 M^2 Q^2-9 Q^4\right)  +3 M Q^4\right) = 4 r_B^4 Q^2+9 r_B^2 Q^4+4 Q^6 \, .
\end{eqnarray}
The above equation is valid at the critical point of phase transition. In fact, eqn. (\ref{davisflat}) should be thought of as the analogue of the relation given by Davis in eqn. (\ref{davis}), but for black holes in flat spacetime in a box. The role of cosmological constant in eqn. (\ref{davis}) is played by the box parameter $r_B$ in eqn. (\ref{davisflat}).  In the absence of a box, eqn. (\ref{davisflat}) gives $M=2/{\sqrt{3}} Q$, which is in conformity with eqn. (\ref{davis}) for $\Lambda=0$. The parameter range in the relation in eqn. (\ref{davisflat}) is constrained by the requirement that the black hole does not encounter a naked singularity~\cite{Davies:1989ey} and is plotted in figure-(\ref{rbQM}).
Further, the point in parameter space where the charged black holes in asymptotically flat spacetime in a box, become critical is~\cite{Carlip:2003ne,Lundgren:2006kt,Ma:2016vop,Kubiznak:2016qmn}:
\bea \label{criticalflat}
\frac{Q_c}{r_B} = (\sqrt{3} - 2), \, \frac{M_c}{r_B} = \frac{2}{\sqrt{5}} (\sqrt{5} - 2)\, ,
\eea
together with the critical value of radius $r_c = r_B \sqrt{5} (\sqrt{5} - 2)$. The values of parameters in eqn. (\ref{criticalflat}) now actually satisfy the new relation presented in eqn. (\ref{davisflat}), valid at the critical point of thermodynamic phase transition. Relations in eqns. (\ref{davisflat}) and (\ref{criticalflat}) allow us to predict that it might be possible to see the values taken by thermodynamic parameters at the critical point, by studying the classification of circular orbits of test particles in the background of charged black holes in flat spacetime, in a box. Such a computation would require putting appropriate boundary conditions for physical quantities and geodesics, such that they fall off appropriately at the boundary of the cavity, possibly using the methods being developed in~\cite{Basu:2016srp,Dias:2018zjg}. In the case of black holes in AdS, it would also be nice to perform a full classification, including the case of naked singularities~\cite{Pugliese:2013xfa} as well. More importantly, it should be explored whether the connection between circular orbits and second order phase transitions of black holes found here, is a mere coincidence (see comments at the end of section-2), specific to AdS type backgrounds, or a more deeper physical picture remains to be uncovered.  These works are in progress.\\

\vskip 0.5cm

\noindent
{\bf Acknowledgements}
\vskip 0.5cm

\noindent
Authors would like to thank the anonymous referee for helpful suggestions.

\appendix
\section{Appendix}

$\epsilon_h$ which appears in Table-2 in the section-2, is the largest positive number of the set $\{\epsilon_{h1}, \epsilon_{h2}, \epsilon_{h3}, \epsilon_{h4}\}$ described below:
\begin{equation}
    \epsilon_{h1,2} = \sqrt{2 k_2^2 - 8 - S(k_1,k_2) \pm \sqrt{-4 S(k_1,k_2) - 2 p(k_1,k_2) + \frac{q_1(k_1,k_2)}{S(k_1,k_2)}}}
\end{equation}
\begin{equation}
    \epsilon_{h3,4} = \sqrt{2 k_2^2 - 8 + S(k_1,k_2) \pm \sqrt{-4 S(k_1,k_2) - 2 p(k_1,k_2) - \frac{q_1(k_1,k_2)}{S(k_1,k_2)}}}
\end{equation}
where the functions $p(k_1,k_2)$,$q_1(k_1,k_2)$, and $S(k_1,k_2)$ are defined as follows:
\begin{equation}
    p(k_1,k_2) = 6912 k_1^2 k_2^6 - 192 k_2^4 + 1376 k_2^2 - 2688 \nonumber
\end{equation}
\begin{equation}
    q_1(k_1,k_2) = 12800 k_1^2 k_2^4 - 27648 k_1^4 k_2^2 - 18432 k_1^2 k_2^2 \nonumber
\end{equation}
\begin{equation}
    S(k_1,k_2) = \frac{1}{2} \sqrt{-\frac{2 p(k_1,k_2)}{3} + \frac{1}{3}\bigg(q_2(k_1,k_2) + \frac{\Delta_0(k_1,k_2)}{q_2(k_1,k_2)}\bigg)} \nonumber
\end{equation}
\begin{equation}
    q_2(k_1,k_2) = \sqrt[3]{\frac{\Delta_1(k_1,k_2) + \sqrt{\Delta_1(k_1,k_2)^2 - 4 \Delta_0(k_1,k_2)^3}}{2}} \nonumber
\end{equation}
\begin{eqnarray}
    \Delta_0(k_1,k_2) &&=  248832 k_1^4 k_2^2 + 4096 k_2^4 - 18432 k_1^2 k_2^4 \\ \nonumber
    && + 103680 k_1^2 k_2^4 - 2048 k_2^6 + 10752 k_1^2 k_2^6 + 256 k_2^8 \nonumber
\end{eqnarray}
\begin{eqnarray}
    \Delta_1(k_1,k_2) &&= \\ \nonumber
    && 8192 k_2^4 \left(
    39366 k_1^8 + k_2^2(k_2^2-4)^3 \right. \\ \nonumber
    && \left. + 729 k_1^6 (13 k_2^2 - 36) + 9 k_1^2 k_2^2 (72 k_2^4 - 40 k_2^2 + 48) \right. \\ \nonumber
    && \left. + 27 k_1^4 (47 k_2^4 - 72 k_2^2 + 432)
    \right) \nonumber
\end{eqnarray}
and $k_1=l/Q, k_2= M/Q$.




\end{document}